\documentclass[sigconf,screen]{acmart}
\usepackage{enumitem}
\usepackage{subcaption}
\usepackage{xspace}
\usepackage[frozencache]{minted}
\usepackage{soul}
\usepackage{wrapfig}
\usepackage{ifthen}
\DeclareEmphSequence{\bfseries}
\graphicspath{{./figures}}
\usepackage{enumitem}
\setlist{nosep}

\let\oldquote\quote
\let\oldendquote\endquote

\renewenvironment{quote}[1][]{
\def\speakername{#1}%
\oldquote
\itshape
``}{''
\normalshape
\hfill\ifthenelse{\equal{\speakername}{}}{}{(\speakername)}
\oldendquote
}
\newcommand{\inline}[2][]{
\textit{``#2''}\ifthenelse{\equal{#1}{}}{}{(#1)}
}

\DeclareRobustCommand{\projectname}{\textsc{UXAgent}\xspace}

\newmintinline[code]{html}{}

\copyrightyear{2025}
\acmYear{2025}
\setcopyright{rightsretained}
\acmConference[CHI EA '25]{Extended Abstracts of the CHI Conference on
Human Factors in Computing Systems}{April 26-May 1, 2025}{Yokohama, Japan}
\acmBooktitle{Extended Abstracts of the CHI Conference on Human Factors in
Computing Systems (CHI EA '25), April 26-May 1, 2025, Yokohama,
Japan}\acmDOI{10.1145/3706599.3719729}
\acmISBN{979-8-4007-1395-8/2025/04}

\begin{document}

\title{\projectname: An LLM-Agent-Based Usability Testing Framework for Web Design}

\author{Yuxuan Lu}
\authornote{This work was done when Yuxuan was an intern, and Dakuo was a visiting scholar at Amazon. Contact email: \{\href{mailto:lu.yuxuan@northeastern.edu}{lu.yuxuan}, \href{mailto:d.wang@northeastern.edu}{d.wang}\}@northeastern.edu
}

\affiliation{%
  \institution{Northeastern University}
  \country{USA}
}

\author{Bingsheng Yao}
\affiliation{%
  \institution{Northeastern University}
  \country{USA}
}

\author{Hansu Gu}
\affiliation{%
  \institution{Amazon}
  \country{USA}
}

\author{Jing Huang}
\affiliation{%
  \institution{Amazon}
  \country{USA}
}

\author{Zheshen (Jessie) Wang}
\affiliation{%
  \institution{Amazon}
  \country{USA}
}

\author{Yang Li}
\affiliation{%
  \institution{Amazon}
  \country{USA}
}

\author{Jiri Gesi}
\affiliation{%
  \institution{Amazon}
  \country{USA}
}

\author{Qi He}
\affiliation{%
  \institution{Amazon}
  \country{USA}
}

\author{Toby Jia-Jun Li}
\affiliation{%
  \institution{University of Notre Dame}
  \country{USA}
}

\author{Dakuo Wang}
\affiliation{%
  \institution{Northeastern University}
  \country{USA}
}

\renewcommand{\shortauthors}{Lu et al.}

\begin{abstract}
Usability testing is a fundamental yet challenging  research method for user experience (UX) researchers to evaluate a web design.
Recent advances in Large Language Model-simulated Agent (\textbf{LLM Agent}) research inspired us to design \projectname to support UX researchers in evaluating and reiterating their usability testing study design before they conduct the real human-subject study.
Our system features an LLM Agent module and a universal browser connector module so that UX researchers can automatically generate thousands of simulated users to test the target website.
The system can generate UX study results in qualitative (e.g., interviewing how an agent thinks), quantitative (e.g., \# of actions), and video recording formats for UX researchers to analyze. 
Through a heuristic user evaluation with five UX researchers, participants praised the innovation of our system but also expressed concerns about the future of UX study with LLM Agents\footnote{Our demo and code is available at \url{https://uxagent.hailab.io}}.
\end{abstract}


\begin{teaserfigure}
    \centering
    \includegraphics[width=1\linewidth]{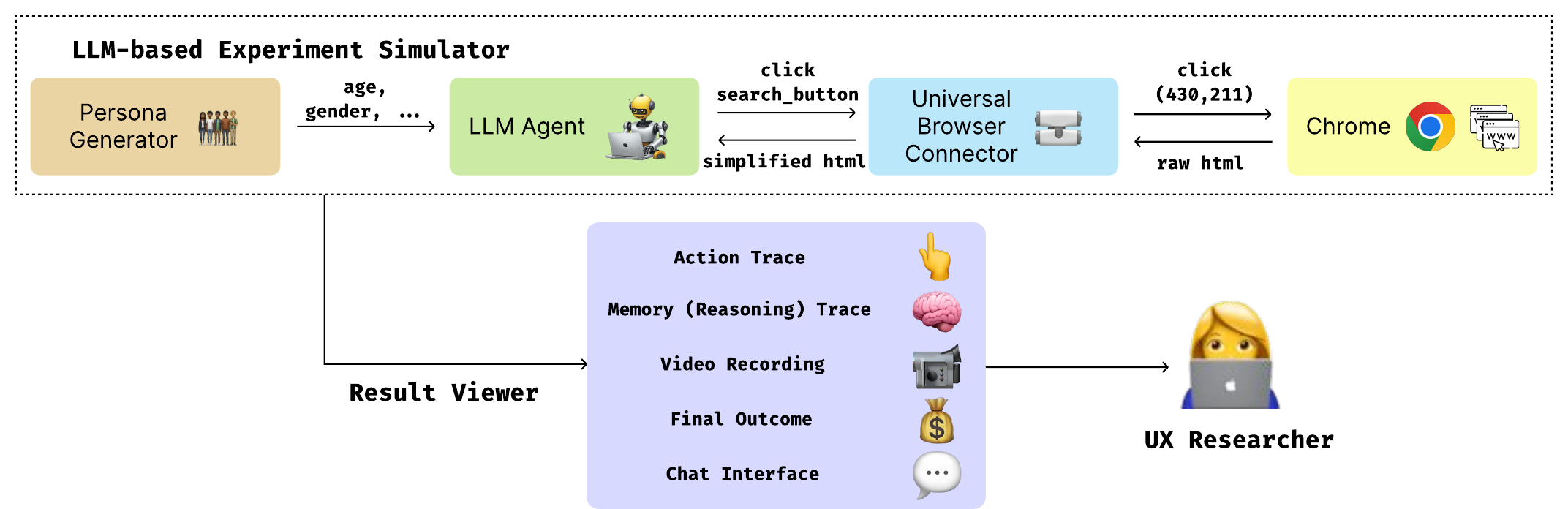}
    \caption{UXAgent System Design. An UX researcher can use the persona generator to generate thousands of user personas and feed them into the LLM Agent. The LLM Agent interacts with the target web design through a Universal Browser Connector. Collected multimodal data are presented to the UX researcher through a Result Viewer.}
    \Description{Diagram illustrating the UXAgent System Design. It consists of multiple interconnected components: (1) A 'Persona Generator' creates user personas with attributes like age and gender. (2) The 'LLM Agent' receives these personas and interacts with web designs through a 'Universal Browser Connector,' which processes simplified and raw HTML interactions. (3) The browser connector communicates with 'Chrome,' executing actions like clicks at specified coordinates. (4) The collected multimodal data, including action traces, memory/reasoning traces, video recordings, final outcomes, and chat interactions, are displayed in a 'Result Viewer.' (5) Finally, a 'UX Researcher' analyzes the presented data.}
    \label{fig:system-arch}
\end{teaserfigure}

\maketitle

\section{Introduction}

Usability testing is a fundamental research method for evaluating the user experience (UX) of a new feature or a new web page design~\cite{shawHandbookUsabilityTesting1996a}.
Typically, the UX researcher (or a UI/UX designer, product manager, etc.) leads the design and execution of a usability testing study: they start with a designed new feature, and they need to come up with an experiment design (e.g. A/B testing and the user tasks), recruit participants, execute the study, collect, and analyze the data.
However, the current process of usability testing often faces challenges, especially in the experiment design stage and the participant recruitment stage \cite{folstadAnalysisPracticalUsability2012, hertzumEvaluatorEffectChilling2003, kuangMergingResultsNo2022, norgaardWhatUsabilityEvaluators2006}.
In the experiment design stage, it is hard for UX researchers to~\textbf{evaluate and gather early feedback on their experiment design}.
After days (or weeks) of experiments, an ill-designed experiment design
cannot yield useful
results 
to support the feature design iteration~\cite{shawHandbookUsabilityTesting1996a}.
Another common challenge is that UX researchers often find it \textbf{difficult to recruit enough qualified participants}, especially when the target users are from a narrowly defined user group.
We all acknowledge that an ill-designed experiment is irresponsible to human participants, as it ultimately wastes their valuable time.
But how can we help UX researchers gather early feedback on their study design so that they can iterate it to be more responsible for the human participants?

With the advancement of Large-Language Models (LLMs), some researchers have prompted LLMs to act as autonomous agents (i.e., \textbf{LLM Agent}) that can perceive the environment and generate human-like actions to interact with the environment \cite{wangSurveyLargeLanguage2024}.
LLM Agents have been successfully used in various scenarios, e.g. simulating a group of 25 agents living in a village \cite{parkGenerativeAgentsInteractive2023},
simulating participants in a social science study \cite{parkGenerativeAgentSimulations2024,schmidgallAgentClinicMultimodalAgent2024,leeApplicationsGPTPolitical,gurcanLLMAugmentedAgentBasedModelling2024}, acting as patients and clinicians in a hospital \cite{liAgentHospitalSimulacrum2024}, and acting as employees of a software developing company \cite{qianChatDevCommunicativeAgents2024}.
The most notable effort is that \citet{parkGenerativeAgentsInteractive2023} illustrated a simulated town with 25 LLM Agents, who can interact with each other and with the surrounding environment; this work argues that LLM Agents can generate ``believable'' human-like behaviors and reasonings, where ``believable'' implies that the agents ``appear to make decisions and act on their own volition'' like a real human~\cite{parkGenerativeAgentsInteractive2023, chen2025towards}.
For instance, WebAgent~\cite{gurRealWorldWebAgentPlanning2023} and Claude's \textit{Computer-Use} API\footnote{https://www.anthropic.com/news/3-5-models-and-computer-use} are two recent examples where researchers allowed the model to interact with external environments to complete tasks such as information retrieval and question answering.
Although these works enable LLMs to interact with external environments (e.g., a webpage), they do not focus on the simulation of diverse user behaviors like real human users. 
For example, in the scenario of online shopping,  these LLM Agents are optimized to make the final purchase using the most optimized path with the least action steps, while human users often have a twisted shopping path with lots of seemingly wasted action steps along the way.

Our project explores the possibility of utilizing LLM Agents' capabilities (i.e., simulating human behaviors and reasoning processes) to support UX researchers for web design.
The usability testing guideline\cite{barnum2020usability} poses the following design requirements for our system: 1) the system should be able to interact with any web environment with minimal or no customization (\textbf{easy-to-use and generalization}), 2) the system should be able to produce \textbf{both quantitative and qualitative usability data} (e.g., interviews, questionnaires, or even video recordings) so that the researcher can use their familiar analysis methods, and 3) the system should balance \textbf{group representativeness and individual uniqueness} (e.g., the female LLM Agent group and a single agent named Mary Jane, 35-year-old female, lives in Boston, etc.).

Based on these requirements, we propose \projectname, a system that can generate LLM Agents as usability testing participants at scale and run simulated interactions with a given web environment to collect simulated user behavioral data.
The UX researcher can define a demographic distribution and use the Persona Generator to generate thousands of personas for the LLM Agents,  and then use the Universal Browser Connector module to interact with webpages via Chrome, which generates large-scale user behavior traces. 
The Universal Browser Connector module allows LLM Agents to seamlessly parse and interpret web pages while automatically executing actions on the webpage without the need for manually predefined action spaces.
The system produces video recordings, as well as features a chat interface to allow the UX researchers to view the LLM Agents' interactions with the webpage and conduct qualitative interviews in natural language conversation , empowering the UX researcher to collect qualitative data. 
Our position is that: \textbf{LLM Agents are not to replace human participants, rather to be more responsible to the human participants --- LLM Agents can work together with UX researchers (human-AI collaboration~\cite{wangHumanAICollaborationData2019}) in a simulated pilot session to provide the desired early and immediate feedback for UX researchers to further iterate the design of the human-subject study before conducting it.}

To evaluate our system, we further conducted a user study with 5 UX researchers and asked them to examine the \projectname-generated data (video recording,  action trace,  memory log, and the final outcome) of how 60 LLM Agents interact with a shopping website.
The user study results suggested that our  
UX researcher participants felt that the generated human behavior data are ``not like real humans'' because they are ``very detailed'' and ``real human [users] won't think like that'', but they still find the data generated from our system ``very helpful'', which can indeed help them iterate their experiment design. 
We conclude our paper with discussions of how LLM Agents may alter the future of UX research.

\section{Related Work}
\subsection{Challenges in Usability Testing}
Usability testing is a core component of UX research, used to evaluate how easily users can interact with a product to achieve their goals \cite{shawHandbookUsabilityTesting1996a,barnum2020usability, bastienUsabilityTestingReview2010, lewis2012usability}.
It involves observing real users as they navigate through tasks and providing valuable feedback on design effectiveness \cite{barnum2020usability}. 
This method helps refine products by identifying usability issues, measuring user satisfaction, and improving overall user experience. The key benefits of usability testing include validating design choices, avoiding internal biases, and ensuring the product meets user expectations.

However, UX researchers conducting usability testing on web designs have been facing multiple challenges in the experiment design stage and the participant recruitment stage~\cite{folstadAnalysisPracticalUsability2012, hertzumEvaluatorEffectChilling2003, kuangMergingResultsNo2022, norgaardWhatUsabilityEvaluators2006}.
To mitigate these challenges, researchers have begun to explore using agents to simulate usability testing \cite{ren2014agent}.
Prior works in agent-based tools mainly focused on pre-defined environments, like GUI \cite{eskonenAutomatingGUITesting2020} and games \cite{stahlkeArtificialPlayfulnessTool2019,fernandesAgentsAutomatedUser2021}.
Nevertheless, the potential for employing agents in web design usability testing remains underexplored, largely due to the constraints of traditional web automation technologies.

\subsection{LLM Agent and LLM Web Agent}

In the field of HCI, recent studies have demonstrated that LLMs have the capability to simulate human-like behaviors, making research on personalized agent behavior feasible. Unlike task-oriented agents, which are typically focused on completing specific tasks, these human agents or role-play agents \cite{chen2025towards} are designed with diverse personas that reflect not only their roles and expertise but also their preferences and habits. For example, \citet{parkGenerativeAgentsInteractive2023} developed a simulated town with 25 LLM Agents, each with a unique persona, and exhibited believable human behaviors through their interactions. 
LLM Agents are also used to study users' privacy concerns \cite{zhangPrivacyLeakageOvershadowed2025,chenEmpathyBasedSandboxApproach2024}. For example, \citet{chenEmpathyBasedSandboxApproach2024} introduced a privacy sandbox where users can modify an agent’s persona, allowing it to generate search histories and profiles based on personal traits.
\citet{taebAXNavReplayingAccessibility2024} proposed AXNav that leverages LLM to convert manual accessibility test instructions to replayable and navigatable videos.
These innovations guide our work, as we aim to simulate user behavior in web environments by incorporating diverse agent personas for usability testing.

LLM Agents in web environments (LLM Web Agents) like WebGPT \cite{nakanoWebGPTBrowserassistedQuestionanswering2022} and LASER \cite{maLASERLLMAgent2024} have also demonstrated that LLM Agents can perform more complex tasks within web environments, such as searching, browsing and shopping \cite{yaoWebShopScalableRealWorld2022}.
With these advancements, LLM Web Agents present a promising solution for simulating human subjects in usability testing while interacting with web environments.
To this end, we propose \projectname, an LLM-Agent-based system for automating usability testing in web environments.

\section{\projectname: LLM-Agent-Based System for Usability Testing}

We propose \projectname, a system that can generate LLM Agents as usability testing participants at scale and run simulated interactions with a given web environment to collect simulated user behavioral data.
The architecture of the \projectname{} system is depicted in Figure~\ref{fig:system-arch}. The system comprises several key components:
The \textbf{Persona Generator Module} generates a diverse set of large-scale personas, which are provided as input to the \textbf{LLM Agent}. 
The \textbf{LLM Agent} interacts with a Chrome browser through the \textbf{Universal Browser Connector Module}. 
This module parses the raw HTML extracted from Chrome and simplifies it for the agent. 
The agent outputs actions, such as \texttt{click on search\_button}, which the Universal Browser Connector translates into raw actions like clicking on a specific coordinate. 
HTML, agents' actions and memories traces, are collected for further analysis.
The \textbf{Chat Interface}, which can be loaded with an agent's memory trace, is used to provide more qualitative feedback on the web design.

\subsection{Persona Generator Module Design}
To support large-scale simulation of diverse user backgrounds, our system enables the generation of diverse agent personas.
The user can specify a distribution of agent demographics and an example persona; the persona generator module will automatically generate the desired number of personas with random demographic information matching the specified distribution.
To ensure diversity in generated personas, each time we prompt the LLM to generate personas, a randomly selected persona is used as an example.
A hand-crafted persona is used as the initial seed. The prompt used in our system is detailed in Section \ref{sec:prompt-persona}.

\subsection{LLM Agent Designed for Web Environment Interactions}
We designed the LLM Agent that features a \textbf{Memory Stream} and a two-loop structure, inspired by \citet{kahneman2011thinking}.
The \textbf{Fast Loop} enables real-time interaction with the web environment, while the \textbf{Slow Loop} facilitates in-depth reasoning. 
Each loop contains modules, which take inputs such as the agent persona, memories retrieved from the memory stream and observations of the environment and produce outputs such as memories and actions. 
Figure \ref{fig:structure} illustrates the agent’s structure.

The \textbf{Fast Loop} is a fast and responsive loop in which the agent quickly operates the web environment without much in-depth thinking and has a relatively low latency between actions.
It consists of the following modules:
1) Perception Module: The agent makes an observation of the web environment and produces a stream of observation memories in natural language.
2) Planning Module: The agent plans the next step based on the current state of the web environment and the recent memories.
3) Action Module: The agent takes the action on the web environment based on the plan.

The \textbf{Slow Loop} of our agent is designed to provide high-level insights and strategic guidance to the Fast Loop, ensuring that the agent remains aligned with its intent and persona.
It consists of the following modules:
1) Wonder Module: The agent generates random thoughts based on the current situation, mimicking a human's ``mind drifting'' phenomenon.
2) Reflect Module: The agent reflects on recent memories and generates high-level insights and reasoning that will be used to guide their future actions.

The \textbf{Memory Stream} is used to store and retrieve the agent's memories, including its observations, actions, reflections, and thoughts. Each memory piece consists of a sentence or paragraph in natural language combined with a timestamp. 
The Memory Stream also serves as a bridge between the Fast and Slow loops.
Following prior work \cite{parkGenerativeAgentsInteractive2023}, we design our memory retrieval module to focus on importance, relevance, and recency. Each memory is scored based on these factors, with weights tailored to prioritize recency in the Fast Loop for quick responses and relevance in the Slow Loop for deeper reasoning, aligning with human cognitive patterns.

\subsection{Universal Browser Connector Module}

To seek the balance between the complexity of the observation space and the flexibility of the action space, to allow our LLM Agent to operate web browsers automatically, we implemented a Universal Browser Connector that \textbf{uses manually crafted policies to parse the raw HTML to a simplified version of HTML as the model's observation space and use auto-generated action space that can be easily generalized to different websites}.
The simplified HTML is a tree structure that contains the visible information of the web page, such as the title, the product list, the product detail, and the shopping cart, while trimming all unnecessary information, such as the CSS and JavaScript code.
Action space is a set of task-agnostic actions generated by the parser that the agent can take, such as ``click'', ``type'' and ``back'', that can be easily generalized to different websites.
The agent can take actions in the action space to interact with the web environment and observe the web environment by receiving the simplified HTML structure.
The definition of the action space, observation space, and the HTML parser can be found in appendix \ref{sec:html-parser}.

\subsection{Chat Interface}

After the simulation session, the simulated action traces and memory traces will be loaded into the Chat Interface, allowing the UX researcher to interview the agent and collect qualitative feedback. 
Users can first select a persona from the persona selection page and the memory of the persona will be loaded to the chat interface. 
Then, in the chat page shown in Figure \ref{fig:chat-interface-2}, the user can talk and interview the agent to collect qualitative feedback. We also allow uploading images during the chat, allowing the user to get feedback on a new feature prototype, even before it is implemented.

\subsection{Simulation Environment and Task for LLM Agents}

To validate the effectiveness of \projectname, we deployed the LLM Agents on two online shopping platforms: WebArena \cite{zhouWebArenaRealisticWeb2024} and Google Flights\footnote{https://www.google.com/travel/flights}.
WebArena is a publicly available open-source online shopping environment and is used as the target experiment.
We chose WebArena as our experiment platform because it works similarly to Amazon.com and is widely used by various research projects \cite{suLearnbyinteractDataCentricFramework2024, yangAgentOccamSimpleStrong2024}.
Google Flights, a commercial platform for flight booking, is also used to demonstrate the adaptability of our system from the online shopping use case to the travel booking use case.
As shopping task represents a common, highly personal activity that humans engage in daily and allows the agent to exhibit diverse behaviors,
we selected the shopping task (``buy a jacket'') as the agents' initial intent in our study.
From the initial intent, our system is able to generate simulation results, including the agents' memory and action traces, video recording, and final outcome (such as item purchased or session quit).

In our study,  we used the Persona Generator Module to generate 60 agent personas, following a uniform distribution across different gender groups (male, female, and non-binary) and across different income groups (\$0-\$30k, \$30k-\$58k, \$58k-\$94k, \$94k-\$153k, and \$153k-).
We then ran one simulation session with each of these 60 agent personas using \projectname on WebArena.
The simulation results revealed different purchase behaviors across income groups, with the average purchased amount increasing with income: \$28.41 for the \$0-\$30k group, \$15.99 for \$30k-\$58k, \$54.85 for \$58k-\$94k, \$41.03 for \$94k-\$153k, and \$75.34 for \$153k+.
We provided the simulation result for the UX researcher participants to analyze as if they were running these user study sessions.

\section{User Study}

\begin{figure*}[t]
    \centering
    \begin{subfigure}[b]{0.5\linewidth}
        \centering
        \vfill
        \includegraphics[width=1\linewidth]{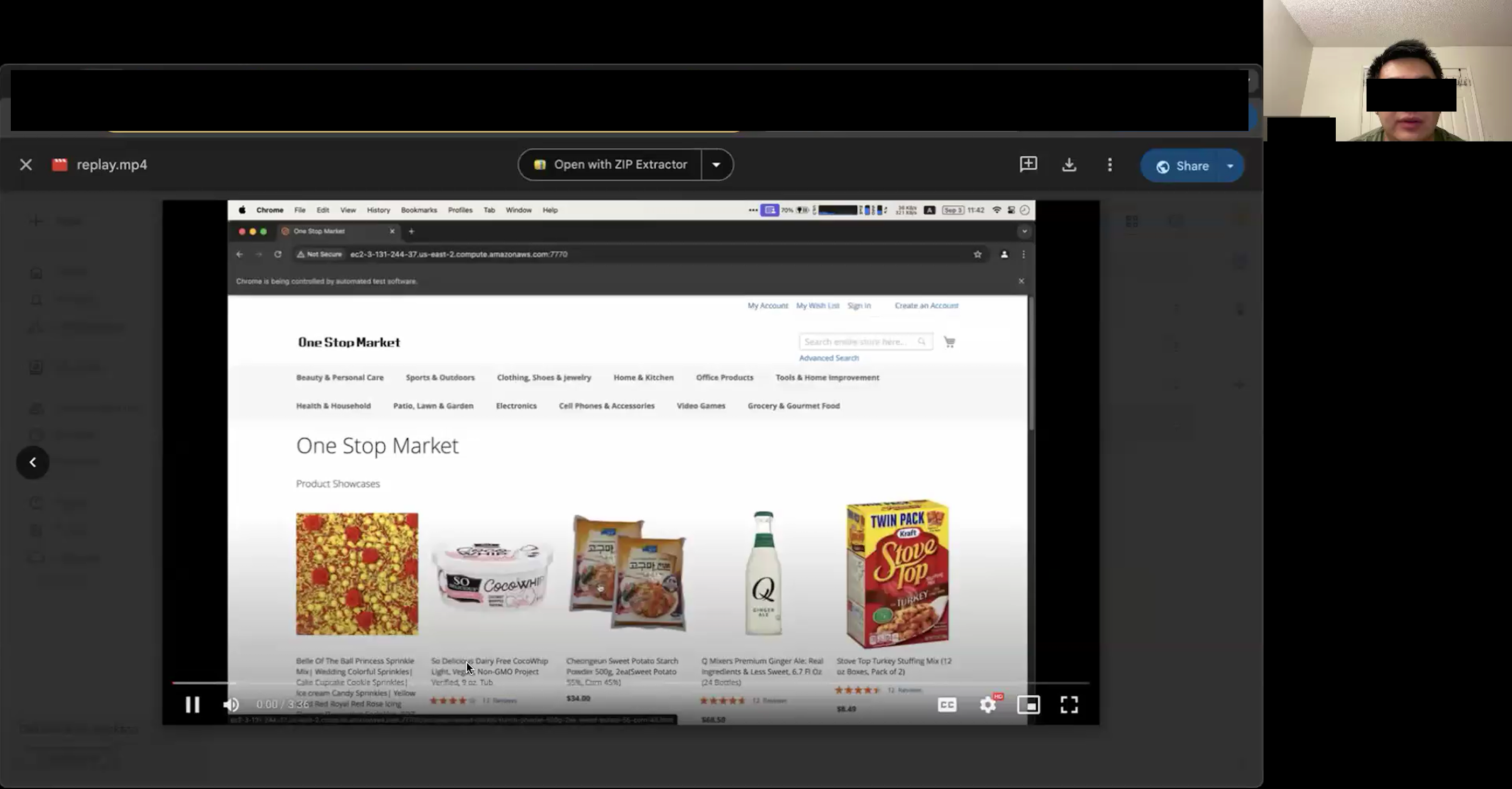}
        \vfill
        
        \vspace{\baselineskip}
        \caption{
        A participant in our user study is watching a video recording replaying an LLM Agent's onling shoping action trace. 
        }
        \label{fig:study-zoom}
    \end{subfigure}
    \hfill
    \begin{subfigure}[b]{0.4\linewidth}
        \centering
        \includegraphics[width=.9\linewidth]{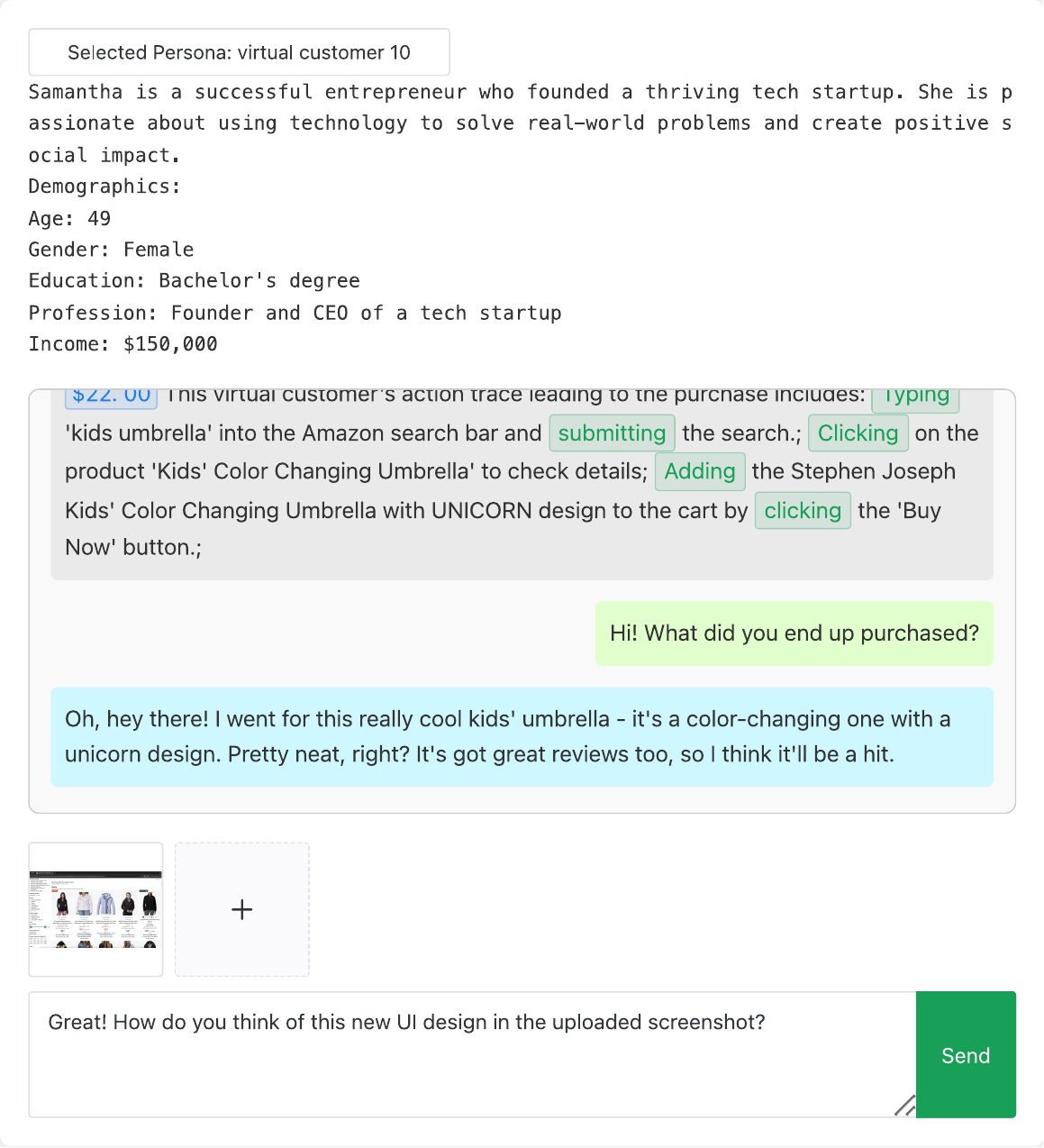}
        \caption{Chat UI. The UX researcher can select a LLM Agent to ask why it had certain behaviors.}
        \label{fig:chat-interface-2}
    \end{subfigure}
    \caption{User study session and an interactive chat interface with LLM Agent}
    \label{fig:study-screenshot}
    \Description{Figure 2 illustrates a user study session and an interactive chat interface with an LLM Agent.

(a) On the left, a participant is watching a video recording replaying an LLM Agent’s online shopping action trace. The screen displays a recorded browsing session on an e-commerce website, where the LLM Agent interacts with various product listings. The participant’s face is partially visible but anonymized with a black bar.

(b) On the right, a chat UI allows a UX researcher to query the LLM Agent about its behavior. The chat window displays a selected persona named 'Samantha,' who is described as a successful entrepreneur with specific demographic details. The agent’s action trace includes searching for a 'kids' umbrella,' clicking on a product, adding it to the cart, and completing the purchase. The UX researcher sends a message inquiring about the purchase decision, and the agent responds with an explanation, mentioning product features and reviews. Below the chat, an option is available for the researcher to upload and receive feedback on UI designs.

This figure highlights how UX researchers can analyze and interact with an LLM Agent’s behavior through video replays and a conversational interface.}
\end{figure*}
\subsection{Participant Recruitment and Participant Task}
We recruited five UX researchers as participants, each with self-reported UX experience ranging from 1 to 6 years (M = 3, SD = 1.87). Most participants rated themselves as ``very familiar'' with usability analysis on a 5-point scale (M = 4.2, SD = 0.83). 
All participants have prior experience using Large Language Models (e.g., ChatGPT, Claude). The study was conducted remotely, with participants using their computers to access the LLM Agent and communicating with the moderator via Zoom. 

Before the study, the moderator used \projectname to generate various personas and asked each agent to finish a simulation session of usability testing on the shopping task.
The moderator then randomly sampled 60 simulation sessions from the 1,000 simulations.
Examples of the generated data are shown in appendix \ref{sec:example-data}.

In our study, participants are instructed to first finish a pre-study survey and then receive a brief introduction to the study tasks.
In the 40-minute study, participants were tasked with conducting a user study to measure customer shopping behavior on a website. Specifically, they were required to analyze simulation data generated by \projectname, with the goal of improving the design of the user study. 
Participants were instructed to freely navigate through various types of data while thinking aloud.
Participants are then invited to interact with a chatbot, pre-loaded with a simulated participant's memory, to ``interview'' the simulated participant.
Participants were provided with three different types of data generated by \projectname, including 1) agent's action trace, 2) memory trace, and 3) final outcome.
Any questions regarding the tasks or the data from the participants were addressed before proceeding.

After completing the data analysis, participants filled out a Likert scale survey to assess the system's efficiency, trustworthiness, satisfaction, and helpfulness, following methodologies from prior work on LLM-generated data~\cite{kuangCollaborationConversationalAI2023}.
Participants were also asked to answer ``For each type of data generated, is it helpful for designing your study?''
We then conducted a semi-structured post-study interview, asking participants to share their perceptions of the different data types and how the simulated results could provide insights on the design of future user studies. We also gathered suggestions for system improvements. 
A screenshot of our user study can be found in Figure \ref{fig:study-zoom}.
Each session lasted approximately 40 minutes, was video-recorded, and the transcriptions were analyzed using an inductive process~\cite{thomasGeneralInductiveApproach2006}.

\subsection{Study Findings}

\subsubsection{How Did UX Researchers Use Our \projectname System?}
Participants followed four categories of analysis paradigm when analyzing data generated by \projectname: 1) gaining trust in the data, 2) making sense of the data, 3) proposing hypotheses and detailed observations, and 4) drawing conclusions.

\paragraph{\textbf{Gaining trust of data}}
When participants had access to all types of data generated by \projectname, their predominant response was to examine each data type to ensure its trustworthiness. 
Most participants only proceeded with deeper analysis after confirming the data's trustworthiness. However, some participants began to analyze the data early on without fully verifying their trustworthiness and encountered challenges at later stages.

\paragraph{\textbf{Making sense of data}}
Once trust was established, they moved on to the sense-making phase, where they tried to interpret the data and understand the relationships between different variables.
For example, participants calculated the average price and number of actions among different groups in a google sheet.
P4 was misled by the persona names presented to them and was trying to analyze the different shopping behaviors of LLM Agents with different names that are randomly generated:
\inline{Most of the Jakes? There's only one Jake here that didn't buy it, but most Jakes here bought it.}
This suggests a better understanding of how the data is generated is necessary for future adoption of such data.

\paragraph{\textbf{Proposing hypotheses and detailed observations}}
Participants then made hypotheses on the experiment result and made detailed observations to verify their hypothesis.
Most participants noticed agents of non-binary genders have a lower purchase rate than male and female agents. P4 hypothesized that maybe this is due to product availability, as there are not many unisex products.
They then checked the raw action log and noticed that a non-binary user searched for a `unisex jacket,' clicked into a `women's jacket,' and ended up purchasing nothing, which proved their hypothesis.

\paragraph{\textbf{Drawing Conclusions}}
Participants then discussed what they found during the analysis of the simulated result.
They organized the findings in a document and discussed how they were going to revise and iterate their study design after the simulation.
P4 suggested that they will move quickly, iterate the study design, and re-run the simulation study to test the new design: \inline{Since it's really quick and easy, I think it could encourage people to do a lot more iterative design.}

\begin{figure*}
    \centering
    \begin{subfigure}[b]{0.48\linewidth}
        \centering
        \includegraphics[width=\linewidth]{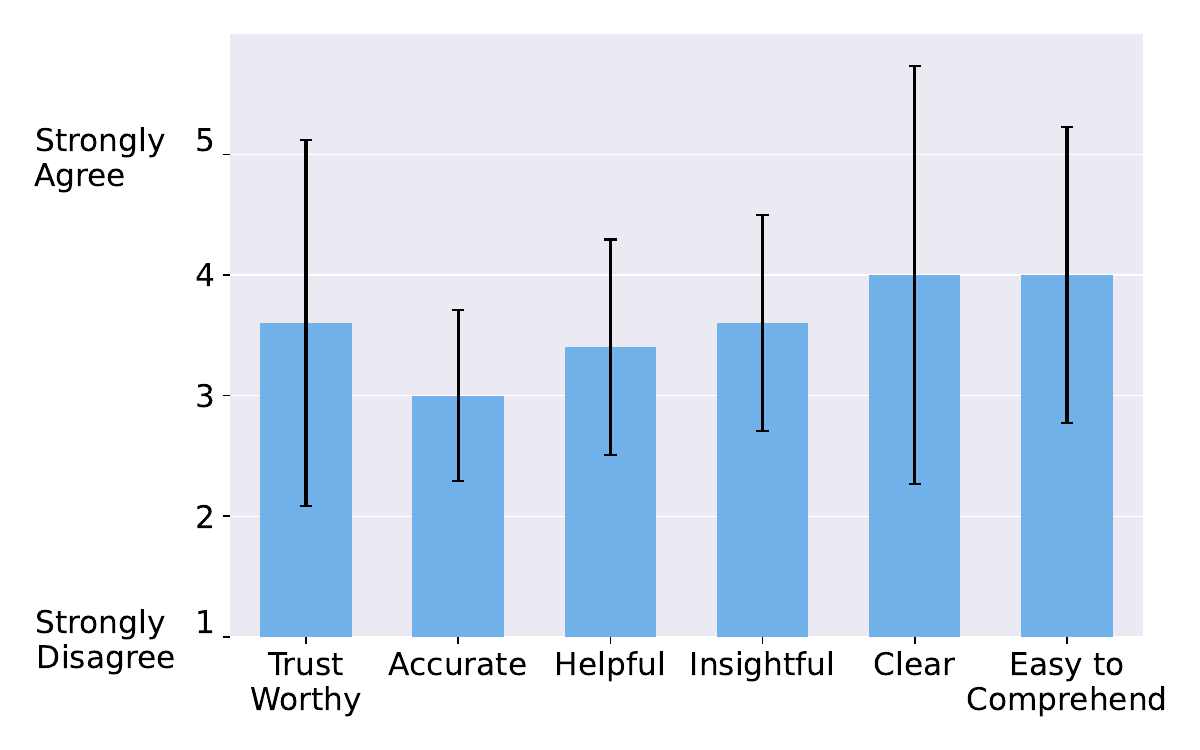}
        \caption{Participant’s Perception towards the Usage of \projectname in User Study}
        \label{fig:overall}
    \end{subfigure}
    \hspace{.02\linewidth}
    \begin{subfigure}[b]{0.48\linewidth}
        \centering
        \includegraphics[width=\linewidth]{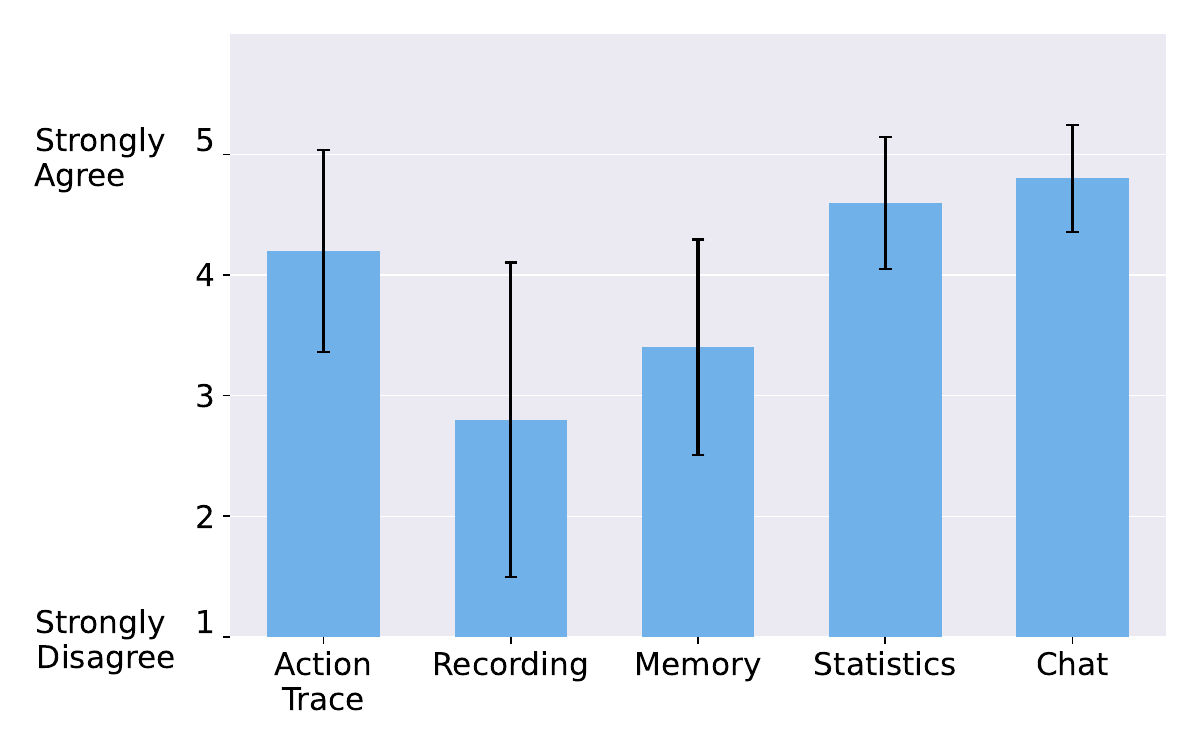}
        \caption{Participant’s 
        Perceived Helpfulness Towards the Different Data Modality Generated by \projectname
        }
        \label{fig:by-category}
    \end{subfigure}
    \caption{Participant's perception towards using \projectname in User Study}
    \label{fig:perception}
    \Description{This figure presents two bar charts summarizing participants' perceptions of using \projectname in a user study, including the raw data.

(a) The left chart visualizes participants' ratings on various aspects of LLM Agents, measured on a 1 to 5 scale (1 - strongly disagree, 5 - strongly agree). The mean ratings and standard deviations for each aspect are as follows:

Trustworthy: 3.6 (SD = 1.5166)
Accurate: 3.0 (SD = 0.7071)
Helpful: 3.4 (SD = 0.8944)
Insightful: 3.6 (SD = 0.8944)
Clear: 4.0 (SD = 1.7321)
Easy to Comprehend: 4.0 (SD = 1.2247)
(b) The right chart depicts participants' perceptions of different types of data generated by \projectname, with the following mean ratings and standard deviations:

Action Trace: 4.2 (SD = 0.8367)
Recording: 2.8 (SD = 1.3038)
Memory: 3.4 (SD = 0.8944)
Statistics: 4.6 (SD = 0.5477)
Chat: 4.8 (SD = 0.4472)
Easy to Comprehend: 4.0 (SD = 1.2247)
This figure illustrates user perceptions regarding the reliability and usefulness of LLM-generated data, with chat and statistics receiving the highest ratings, while recordings scored the lowest."}
\end{figure*}

\subsubsection{Participants' Perceptions and Concerns of \projectname}
\paragraph{\textbf{Participant have mixed feelings of simulated data from \projectname}}
Figure~\ref{fig:perception} shows participants' perceived usage of LLM Agents generated human behavior data. 
Notably, although the generated data may not seem very realistic (M=3, SD=0.7),
participants still found the data presented by \projectname to be very helpful (M=3.4, SD=0.89).
This perceived helpfulness was also seen as an expansion of the pilot user study design, with P1 mentioning:
\begin{quote}[P1]
    It's very hard to find pilot study of 60 participants or even just five or ten. So I think in this way I can conduct my pilot study with the large language model agent.
\end{quote}

Participants believed that the simulation results were insightful (M=3.6, SD=0.89) and trustworthy (M=3.6, SD=1.51), sparking an interest in further investigation—an uncommon occurrence in smaller-scale pilot studies.
Furthermore, the large-scale data allows them to see things from others' perspectives.
Participants especially liked the ability to talk with the agent (M=4.8, SD=0.44) and collect information that you may never be able to get from real human participants, as P4 mentioned: \inline{I think it could help me to kind of brainstorm.}
The final outcome and statistics (M=4.6, SD=0.54) and the agents' action trace (M=4.2, SD=0.84) were also rated as helpful by the participants.

Participants mentioned that the raw memory traces are hard to read and analyze.
P2 mentioned that \inline{If you can present the thoughts in a better way, like not just right now for each agent, probably like a highlight or summary.}
We believe this is due to the noisy nature of all unprocessed raw data, and we further discuss the possible use of this data in Section \ref{sec:discussion-design-considerations}. Participants gained trust in the generated data from watching the generated video recording of the browser operated by the LLM Agent, but they found analyzing it hard and time-consuming.

\paragraph{\textbf{Concerns about model bias and data bias}}
Participants are concerned about the potential biases introduced by the data generated by LLM Agents. 
In our study, we identified two types of bias that are key concerns for participants when it comes to LLMs.
1) \textbf{Data Representation Bias} is where the decisions made by the agents may be influenced by biases present in the real-world data they were trained on, as stated by P2: \inline{because we do feel like the LLM has sort of stereotyping that. My feeling female will buy more male will buy less things}.
2) \textbf{Algorithmic Decision Bias}, which is introduced in the pre-training and fine-tuning stage of the LLM could distract the study design, as P1 commented,\inline[P1]{it could be bias or it could disturb my study design}.

\paragraph{\textbf{Perceived unrealistic human behavior}}
Participants have varied opinions about whether certain types of the LLM's simulated data are realistic. 
Human users usually quickly skim and disregard unimportant webpage elements like ads or irrelevant results. 
Our agent mirrors this behavior by assigning varying ``importance'' scores to each memory in its memory stream.
However, as the importance score is not shown to our participants, participants reported that the memory stream is ``of too much detail'' and ``not realistic''.
Despite the fact that the presented memory stream may not be a full mimic of real human behavior, participants still showed their love in having these data, as these are still something they cannot collect from experiments with real human participants and can provide additional insights.

\section{Discussion}

\subsection{Design Considerations for  LLM Agent Systems in User Study}
\label{sec:discussion-design-considerations}
The shift towards using LLM to assist UX research has demonstrated immense potential.
Our work \projectname demonstrated that LLM Agents can support user studies with faster design iterations through user behavior simulation, reducing costs and improving the quality of study design~\cite{decker-maurerMethodMadnessUsability2012}.
Our participants suggested that the \textbf{future systems should automatically generate high-level insight summaries} as the generated LLM Agents' raw memory is hard to analyze.
Thus, future systems should explore how to automatically utilize the current data to generate high-level insights and summaries for users.

\subsection{Resolving Risk and Privacy Conerns}
While LLM-assisted UX study systems offer substantial benefits for user studies, they also introduce significant risks and concerns, particularly around privacy and ethical use \cite{liHumanCenteredPrivacyResearch2024, shaoPrivacyLensEvaluatingPrivacy2024}. 
A notable concern about sensitive data might arise in applying LLM Agents in user studies in fields such as healthcare \cite{perisPrivacyTimeLanguage2023}.
Our participants viewed data from LLM Agents as supplementary rather than definitive, which is consistent with prior findings that LLMs cannot fully replace human subjects in psychological experiments~\cite{cui2024can}. 
This preference reflects their ability to detect biases, limiting acceptance of LLMs as accurate representations of outcomes~\cite{schmidtSimulatingHumanHCD2024}.

Another critical issue is the risk of researchers misusing AI-generated data \cite{perisPrivacyTimeLanguage2023}. 
Some may be tempted to use this simulated data as a substitute for actual human participant data, potentially bypassing real user studies altogether. 
This could lead to flawed conclusions, as the AI-generated data, while useful for simulations, cannot fully replicate the complexity of real human behavior.
We believe that addressing these risks requires a clear understanding of the limitations of LLM-generated data and establishing rigorous ethical guidelines and robust privacy safeguards to ensure responsible and effective integration of LLM in UX research.

\subsection{Limitations and Future Works}
There are several limitations in our work. 
First, the limited number of participants in our heuristic evaluation may not fully capture the diversity of user experiences or the generalizability of the findings. 
A larger scale of rigorous user testing in the future will help us further explore the effectiveness of \projectname.
Also, our current work focuses only on the qualitative analysis of the participants' perspectives toward LLM Agent simulation, but not on quantitative evaluations of the LLM Agents' versus real humans' shopping behaviors. 
It is crucial to conduct a systematic analysis comparing LLM-generated simulations with real human participants in order to further understand how UX researchers could benefit from using such agents.
Furthermore, our current system and agent design only process semantic or textual information (i.e. HTML) available on the webpage, while visual elements such as images and visual layout information are excluded during parsing. 
Incorporating Multimodal LLMs (MLLMs) to interpret and utilize both textual and visual information could enable the system to better understand page context, enhancing its adaptability in real-world scenarios where visual cues and layout are also critical for effective user interaction modeling.

Our current results confirm the effectiveness of \projectname in web-related user research, as well as the potential of LLMs in simulating user behavior to assist rapid iteration of user research. 
In the future, this work could be expanded to encompass a broader range of user research contexts beyond web applications. 
This might include adaptation to UI/UX design in desktop, mobile, or mixed-reality environments, where LLM Agents could simulate user interactions in complex, multi-modal interfaces. 
Also, we only tested giving the agents an explicit intention of ``buy a jacket''. Experimenting with other intents, such as window shopping, can also help understand the ability and limits of LLM Agent.
Additionally, the action space of the agent is currently limited to basic actions such as ``click'' and ``type''. Supporting other types of actions, such as ``read'', ``scroll'' and ``mouse hover'' can allow the experimenter to collect more information useful to their study.

\section{Conclusion}

In this work, we designed \projectname, a system enabling researchers to conduct simulated user studies, thereby facilitating iterative refinement of their UX study designs.
The results of our evaluation study suggest \projectname can support UX researchers in designing and refining their user studies to improve UX study quality and reduce risk for real human subjects. 
Our study is the first step towards a promising future of designing LLM agents to collaborate with UX researchers to achieve ``human-AI collaboration'' in the field of UX research.

\bibliographystyle{ACM-Reference-Format}
\bibliography{reference,set}

\clearpage
\appendix

\begin{figure*}[t]
    \centering
    \begin{subfigure}[b]{.3\linewidth}
          \centering
          \includegraphics[width=.66\linewidth]{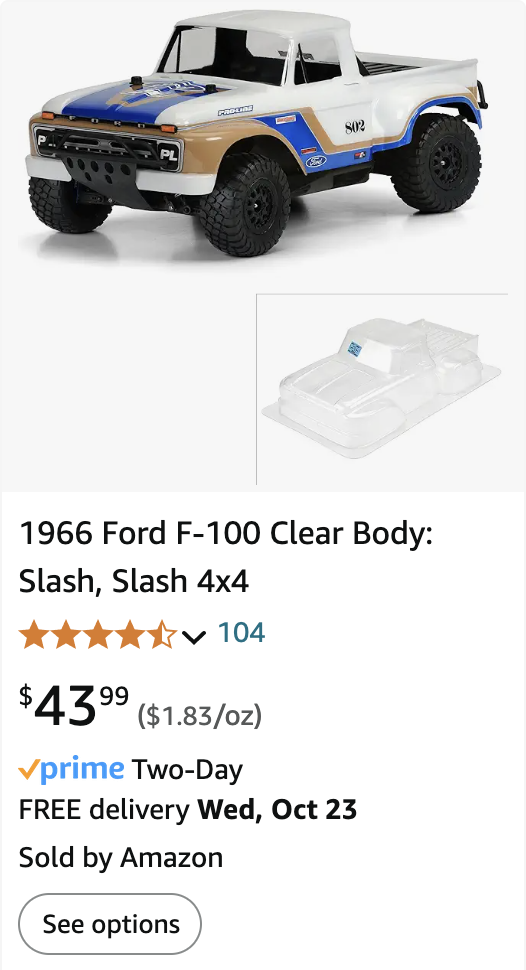}
          \caption{Example of a part of a web page processed by the Perception Module}
    \end{subfigure}\hspace{.07\linewidth}
    \begin{subfigure}[b]{.6\linewidth}
        \centering
        \footnotesize
        \begin{minted}[breaklines,breaksymbol={},breakanywhere]{html}
<div name="search_results.1966_ford_f_100_clear_body_slash_slash_4x4"
class="search-result">
    <a name="search_results.1966_ford_f_100_clear_body_slash_slash_4x4.view_product"
       class="product-name">
        1966 Ford F-100 Clear Body: Slash, Slash 4x4
    </a>
    <div class="product-review">
        <span class="product-rating">4.3 out of 5 stars</span>
        <span class="product-rating-count">103 reviews</span>
    </div>
    <div class="product-price"><span>$43.99</span></div>
    <div class="product-delivery">FREE delivery Mon, Oct 14</div>
</div>
        \end{minted}
        \caption{The simplified HTML structure parsed by the browser environment}
        \label{fig:simp-html}
    \end{subfigure}

    \vspace{1\baselineskip}

    \begin{subfigure}[b]{.8\linewidth}
        \centering
        \begin{minted}[breaklines,breaksymbol={},breakanywhere]{text}
            The search results display a product listing for the '1966 Ford F-100 Clear Body: Slash, Slash 4x4'. This product has a rating of 4.3 out of 5 stars based on 103 reviews and is priced at \$43.99 with free delivery on Monday, October 14.
        \end{minted}
        \caption{The observation memory produced by the Perception Module}
        \label{fig:observation}
    \end{subfigure}
    \caption{Example of a part of a web page processed by the Univerasl Web Connector and the Perception Module.
    The Univerasl Web Connector parses the raw web page (left) into a simplified HTML structure (right), and the Perception Module generates a descriptive observation memory (bottom) for each segment.}
    \label{fig:perception-demo}
    \Description{
    Screenshot of an Amazon product listing for a '1966 Ford F-100 Clear Body: Slash, Slash 4x4.' The product image at the top shows a blue and white remote-controlled (RC) truck body, while a smaller secondary image below displays a clear, unpainted version of the truck body. The listing includes a 4.5-star rating based on 104 customer reviews. The price is displayed as $43.99 ($1.83 per ounce), with a 'Prime' two-day free delivery option available for Wednesday, October 23. The item is sold by Amazon, and there is a 'See options' button at the bottom. The caption indicates that this web page section was processed by the Perception Module.
    }
\end{figure*}

\begin{figure*}
    \centering
    \begin{subfigure}[b]{.8\linewidth}
        \includegraphics[width=\linewidth]{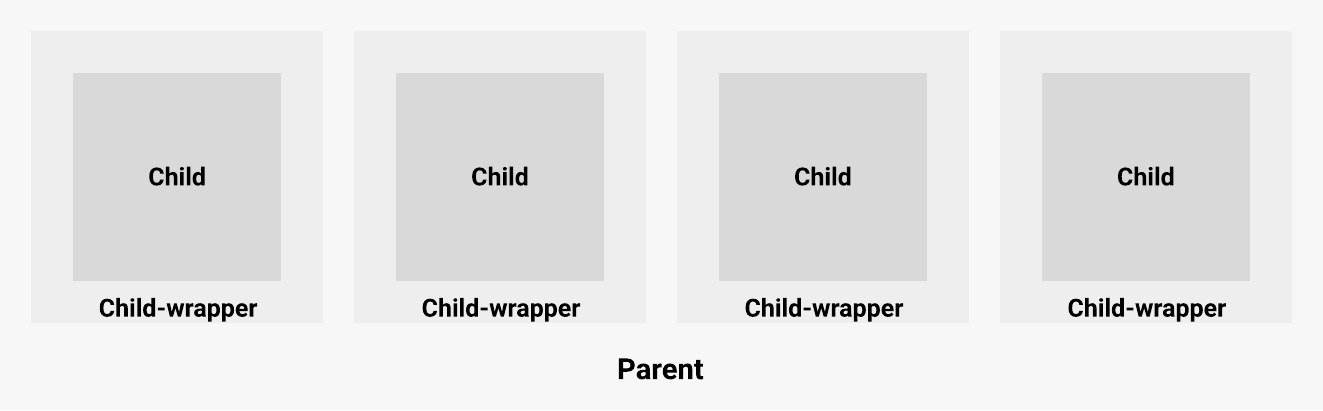}
        \caption{HTML structure before parsing}
        \label{fig:html-parser-example-nested-before}
    \end{subfigure}
    \begin{subfigure}[b]{.8\linewidth}
        \includegraphics[width=\linewidth]{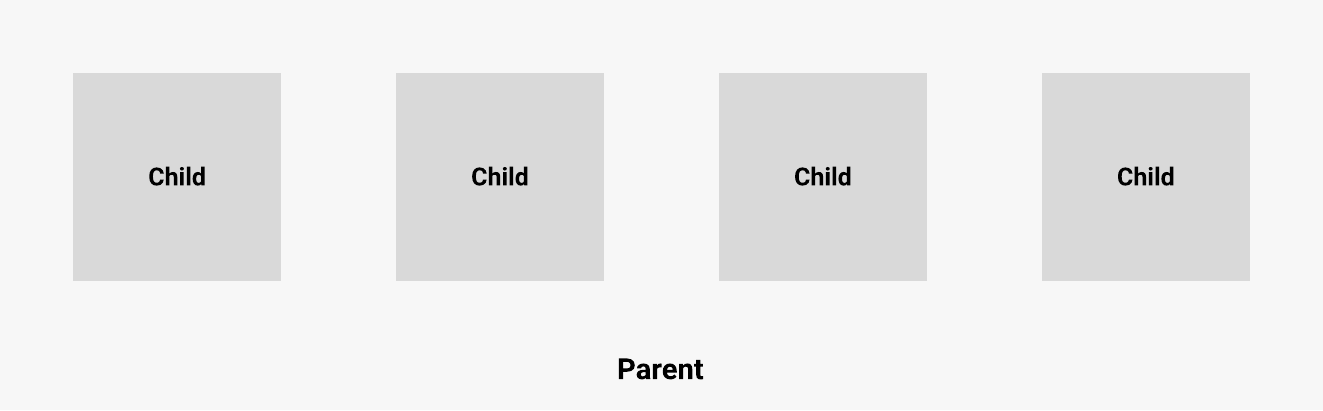}
        \caption{HTML structure after parsing}
        \label{fig:html-parser-example-nested-after}
    \end{subfigure}
    \caption{HTML Parser Example of multiple nested elements.}
    \label{fig:html-parser-example-nested}
    \Description{Figure 5 illustrates an example of parsing an HTML structure with multiple nested elements.

(a) The top diagram represents the HTML structure before parsing. It consists of a 'Parent' element containing multiple 'Child-wrapper' elements, each of which encapsulates a 'Child' element. The child elements are visually enclosed within shaded boxes labeled 'Child,' and each wrapper is labeled 'Child-wrapper.'

(b) The bottom diagram shows the HTML structure after parsing. The intermediate 'Child-wrapper' elements have been removed, leaving only the 'Child' elements directly within the 'Parent' element.

This figure demonstrates how an HTML parser processes nested elements, simplifying the structure by eliminating unnecessary wrapper elements.}
\end{figure*}

\section{Browser Environment}
\label{sec:simulated-browser-environment}

\paragraph{Observation Space}
\label{sec:html-parser}
The observation space in the Universal Browser Connector is represented with a JSON object that contains:
\begin{itemize}
    \item \code{url}: The URL of the current webpage
    \item \code{page}: The simplified HTML of the current webpage
    \item \code{clickables}: A list of clickable elements on the webpage, such as buttons and links
    \item \code{inputs}: A list of input elements on the webpage, such as text fields and dropdowns
    \item \code{error_message}: An error message if an error occurs, for example, if the agent clicked on a button that does not exist
\end{itemize}

To remove redundant information, we only keep the visible information of each HTML tag, such as the text content, the attributes, and child elements.
An example simplified HTML is shown in Figure \ref{fig:perception-demo}.
The parser takes the raw HTML and a recipe and parses the HTML based on the recipe. 
The recipe specifies how to parse the HTML, such as which tags to keep and how to extract the text content and attributes of each tag.
If the recipe has any child elements, the parser will recursively parse the child elements with the child recipe.

\textbf{Handling of Text}
\label{sec:html-parser-text}
Text of an HTML element is extracted if the recipe specifies a \code{add_text} field.
The three possible parse methods of text are \code{default}, \code{text_selector} and \code{text_js}.
\code{default} extracts all of the text content of the tag.
\code{text_selector} extracts the text content of a child element based on a CSS selector.
\code{text_js} serves like a fallback method, where the recipe species a JavaScript function that returns the text content.
In this way, for most cases the recipe can be easily created by selecting the text element. In rare cases, for example, the text of a radio button should be the text of a sibling \code{label} tag, the flexibility of the \code{text_js} method allows the parser to handle these cases.

Sometimes, the visual information and the textual information of an element differ, especially in cases where the HTML's accessibility features are not well-implemented. 
For example, on the product detail page, the review score information is often shown as stars. While an alternative text like ``4.2 out of 5 stars'' exists, there is sometimes no clue that the scores here is the product review.
We allow the recipe to specify a \code{text_format}, for example, ``\code|Rating: {}|'' in this case, to supply the necessary context. With \code{text_format} like this, the abovementioned product review score will be parsed as ``Rating: 4.2 out of 5 stars'', allowing easier understanding by the LLM Agent.

\textbf{Handling of Tag Name and Attributes}
Attributes of HTML elements include 1) the visual information of the element, such as style and class; 2) the semantic information of the element, such as role, href, and src; 3) the information of the element's state, such as checked and selected.
The tag name of an HTML element also sometimes conveys semantic information, such as \code{a} tag is used for hyperlink, and \code{img} tag is used for the image.

In our case, as most visual information is not needed, we remove all visual information attributes from the HTML and keep the semantic information and the state information.
For tag name, the default behavior is to keep the tag name as is, and the recipe can supply a \code{tag_name} field to override the tag name, for example, replace \code{div} with \code{button} for better readability and accessibility.

Specifically, the \code{class} attribute mostly conveys visual information such as color and alignment. However, they sometimes also convey the semantic information, such as \code{product-item-details}.
Thus, the recipe can supply a \code{keep_attr} option to keep the attributes of the tag.
To further enhance flexibility, the recipe can supply a \code{override_attr} option to keep the attributes of the tag based on a JavaScript function.
This way, the parser can handle a wide range of scenarios, from simple attribute extraction to more complex cases where attributes need to be dynamically determined.

The JavaScript state of the HTML element, for example, the \code{selected} attribute and the \code{value} of an input box, is invisible in the HTML, yet important for the agent to know. Thus, for \code{select} and \code{input} elements, we extract their JavaScript state, and add it to the corresponding attribute simplified HTML, for example, a \code{selected} attribute is added to the \code{select} element, and a \code{value} attribute is added to the \code{input} element.

\textbf{Handling of Child Elements}

To develop complex animation to indicate states like hover and focus, UI designers use many nested elements and CSS classes.
For example, a button is often implemented using the \code{button} or \code{div} tag, but to support styles like shadow and animations, web designer often use multiple nested \code{div} element inside the button. While these redundant tags are necessary for better styling, they do not contribute to the semantic information to the button and may affect the agent's understanding.

To simplify the HTML, we ``flatten'' it by removing redundant levels of elements.
For instance, in the HTML shown in Figure \ref{fig:html-parser-example-nested-before}, a nested ``Child-wrapper'' is used. If the recipe specifies a \code{children} field, the parser will select the deep child element (``Child'' in this case) and remove the intermediate ``Child-wrapper''.

\begin{figure}[t]
    \centering
    \begin{subfigure}[b]{.45\linewidth}
        \centering
        \includegraphics[width=.9\linewidth]{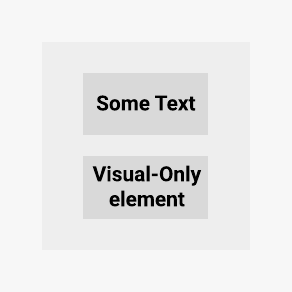}
        \caption{HTML structure before parsing}
        \label{fig:html-parser-example-text-before}
    \end{subfigure}
    \begin{subfigure}[b]{.45\linewidth}
        \centering
        \includegraphics[width=.9\linewidth]{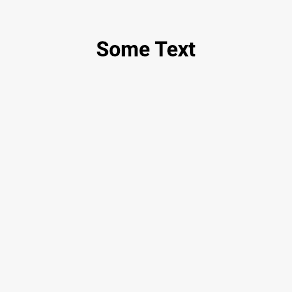}
        \caption{HTML structure after parsing}
        \label{fig:html-parser-example-text-after}
    \end{subfigure}
    \caption{HTML parser example of text in child elements.}
    \label{fig:html-parser-example-text}
    \Description{Figure 7 illustrates an example of an HTML parser processing text in child elements.

(a) The left diagram represents the HTML structure before parsing. It consists of a container element with two nested elements: one labeled 'Some Text' and another labeled 'Visual-Only element.' Both elements are visually enclosed within shaded boxes.

(b) The right diagram shows the HTML structure after parsing. The 'Visual-Only element' has been removed, leaving only the 'Some Text' element inside the container.

This figure demonstrates how an HTML parser processes nested elements by retaining meaningful text content while discarding non-essential visual-only elements.}
\end{figure}

If the recipe does not specify a \code{children} field, all child elements are removed from the HTML. Additionally, as discussed in Section \ref{sec:html-parser-text}, if the recipe specifies an \code{add_text} field, the text from the deep child elements is extracted and added to the corresponding parent tag, which is shown in Figure \ref{fig:html-parser-example-text}. This process reduces the number of tags in the HTML, making it more readable.

In this way, the HTML presented to the LLM Agent is simplified and more readable, while still preserving the same semantic information of the HTML.

\paragraph{Action Space}
\label{sec:action-space}
The action space of the web browser includes the following actions:
\begin{itemize}
    \item \code{click}: perform a mouse click on the target element
    \item \code{type}: type the text into the input box
    \item \code{type_and_submit}: type the text into the input box and submit the form, just like hitting\code{Enter} key after typing the text
    \item \code{clear}: clear the text in the input box
    \item \code{back}: go back to the previous page
    \item \code{terminate}: terminate the current session, denoting user closes the browser
\end{itemize}

For all interactive elements, a \code{name} attribute is appended to the HTML tag to assist the agent in identifying the target element.
The \code{name} attribute is constructed by traversing the parent tree. For instance, if a button is named \code{add_to_cart} in the recipe, and it is a child of a \code{div} named \code{product}, the complete name of the resulting HTML element will be \code{product.add_to_cart}.
Specifically, in our testing online shopping scenario, all the forms only have one input (e.g. search box), thus to reduce the number of action needed and to reduce hullucination, we also provide a composite action \code{type_and_submit}, which is equivlent to typing and hitting the ``enter'' key.  Agents can also \code{type} then \code{click} on the submit button.

\begin{figure}[t] 
    \centering
    \includegraphics[width=\linewidth]{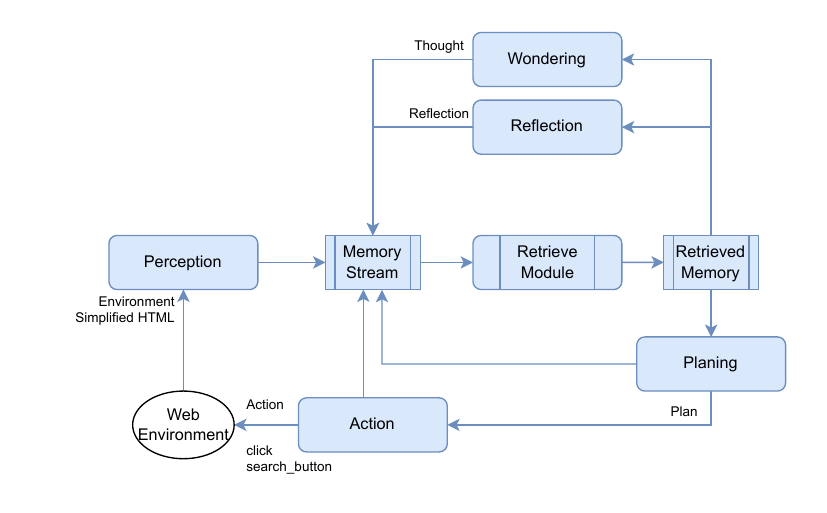}
    \caption{The structure of the LLM Agent's Reasoning Architecture}
    \label{fig:structure}
    \Description{Figure 6 illustrates the structure of the LLM Agent designed for web environment interactions, featuring a Memory Stream and a two-loop architecture: the Fast Loop and the Slow Loop.

The Fast Loop enables real-time interaction with the web environment with minimal latency. It consists of:

Perception Module: Observes the web environment and generates natural language observation memories.
Planning Module: Forms a plan for the next step based on the current state of the web environment and recent memories.
Action Module: Executes the planned action, such as clicking a search button, interacting directly with the web environment.
The Slow Loop provides high-level insights and strategic guidance to the Fast Loop, ensuring alignment with the agent’s intent and persona. It includes:

Wonder Module: Generates random thoughts, mimicking human 'mind drifting.'
Reflect Module: Analyzes recent memories to derive high-level insights and guide future actions.
The Memory Stream serves as a central repository, storing and retrieving the agent’s observations, actions, reflections, and thoughts. Each memory is timestamped and encoded in natural language.

The Retrieve Module fetches relevant memories, prioritizing recency in the Fast Loop for quick responses and relevance in the Slow Loop for deeper reasoning.
Retrieved memories are used in the Planning Module to generate actions, ensuring the agent's reasoning process aligns with human cognitive patterns.
This structure enables the LLM Agent to balance rapid web interactions with thoughtful long-term decision-making, enhancing its adaptability and effectiveness in dynamic online environments.}
\end{figure}

\section{Prompt}
\subsection{Prompt for Persona Generatoin}
\label{sec:prompt-persona}
\begin{minted}[breaklines, breaksymbol={},breakanywhere]{text}
Generate a persona using the above examples. The persona should be different from previous personas to ensure diversity. The persona should:

- Have the age of {age}
- Be {gender}
- Have an income between ${income_range[0]} and ${income_range[1]}

Provide the persona in the same format as the examples.

Only output the persona, no other text.
\end{minted}

\section{Example Data}
\label{sec:example-data}
\subsection{Action Trace}
\begin{minted}[breaklines, breaksymbol={},breakanywhere]{text}
Action 1: type_and_submit, description: Typing 'woman's jacket' into the search input field and submitting the form.
Action 2: click, description: Clicking on the product 'Jackets For Women Womens Hooded Fleece Line Coats Parkas Faux Fur Jackets with Pockets' to view its details.
Action 3: click, description: Clicking on the 'Navy' color option for the jacket.
Action 4: click, description: Clicking on the 'Medium' size option for the jacket.
Action 5: click, description: Clicking on the 'Add to Cart' button to add the chosen product to the cart.
\end{minted}
\subsection{Memory Trace}
This example is filtered for readability.
\begin{minted}[breaklines, breakanywhere,breaksymbol={}]{text}
For action 1, I will: Typing 'woman's jacket' into the search input field and submitting the form.
Before action 2, I thought: I really need to find that navy blue jacket soon. It would be perfect for those networking coffee meet-ups. I hope I can get it on sale.
Before action 2, I thought: If I don't find a job soon, I might have to reconsider my budget for essentials. It feels so overwhelming at times.
Before action 2, I thought: I wonder if I should try making some more connections through social media or if that's just a waste of time. It feels like everything is so competitive right now.
Before action 2, I thought: The action to search for 'woman's jacket' was successful, as the new observation confirms that the search results for this query have been displayed.
Before action 2, I thought: Next, I need to browse through the search results to find a navy blue, medium-sized women's jacket. It will be essential to closely examine the available product options to see if they meet these specific criteria.
Before action 2, I thought: Additionally, I should pay attention to the pricing and consider narrowing down my search to ensure that I am selecting the most affordable option that fits my needs and budget.
Before action 2, I saw: The page title reads 'Search results for: 'woman's jacket''. This provides context for the user about the ongoing search query.
\end{minted}

\end{document}